\documentclass[11pt]{article}
\usepackage{epsfig}

\newcommand{\alps}	{\ensuremath{\alpha_S}}

\newcommand{\df}	{\ensuremath{\mathrm{d}}}
\newcommand{\e}		{\ensuremath{\mathrm{e}}}

\newcommand{\vb}	{\ensuremath{\mathbf{b}}}

\newcommand{\vr}	{\ensuremath{\mathbf{r}}}

\newcommand{\ap}	{\ensuremath{\alpha_{\mathcal{P}}}}
\newcommand{\apm}	{\ensuremath{(\ap-1)}}

\begin{document}
\begin{titlepage}
\renewcommand{\thefootnote}{\fnsymbol{footnote}}
\begin{flushright}
CU-TP-746 \\
Cavendish-HEP-96/04 \\ 
hep-ph/9605302 \\
May 1996
\end{flushright}
\vspace*{10mm}
 
\begin{center}
\textbf{\Large Large multiplicity fluctuations and saturation effects
in onium collisions}
\vspace{10mm}
 
\textbf{A.H.~Mueller}\footnote{This work is supported in part by the US
Department of Energy under Grant DE-FG02-94ER40819} \\
\vspace{3mm}
\textit{Department of Physics, Columbia University,} \\
\textit{New York, NY~10027, USA}\\
\vspace{6mm}

\textbf{G.P.~Salam}\footnote{This work is supported by the UK Particle
Physics and Astronomy Research Council} \\
\vspace{3mm}

\textit{Cavendish Laboratory, Cambridge University,} \\
\textit{Madingley Road, Cambridge CB3 0HE, UK} \\
\vspace{2mm}
e-mail: salam@hep.phy.cam.ac.uk
\end{center}
 
\vspace{20mm}
\begin{abstract}
This paper studies two related questions in high energy onium-onium
scattering: the probability of producing an unusually large number of
particles in a collision, where it is found that the cross section for
producing a central multiplicity proportional to $k$ should decrease
exponentially in $\sqrt{k}$. Secondly, the nature of gluon (dipole)
evolution when dipole densities become so high that saturation effects
due to dipole-dipole interactions become important: measures of
saturation are developed to help understand when saturation becomes
important, and further information is obtained by exploiting changes
of frame, which interchange unitarity and saturation corrections.
\end{abstract}

\end{titlepage}

\section{Introduction}
The Balitsky, Fadin, Kuraev and Lipatov (BFKL) equation
\cite{BaLi78,KuLF77,Lipa86} is meant to
describe the energy dependence of certain high energy hard processes.
For the BFKL equation to be applicable certain conditions must be met:
(i) The process, or part of a process, to be described should have
only a single hard transverse momentum scale.  Processes having more
than one hard scale may have significant contribution from the
Dokshitzer, Gribov, Lipatov, Altarelli and Parisi (DGLAP) equation
\cite{Doks77, GrLi72, AlPa77} which governs scale dependence in QCD.
Certain observables (e.g.\ \cite{MuNa87}) are known to satisfy this
criterion though they have not yet been well measured.  (ii) The
rapidity of the hard process, related to the centre of mass energy $E$
and the hard transverse momentum scale $k$ by $Y = \ln E^2/k^2$,
should be large.  The BFKL equation includes all terms $(\alpha_sY)^n$
in the coupling but neglects factors of $\alpha_s$ not accompanied by
factors of $Y$.  There is a vigorous effort \cite{FaLi96} to
systematically include the first non-leading terms, an $\alpha_s$
unaccompanied by a $Y$ and this should improve the accuracy and
reliability of the BFKL equation.

The BFKL equation should describe high energy processes having a
single hard scale in a limited range of rapidity.  When $Y$ becomes
too large, unitarity corrections become important and slow the rate of
growth with energy.  At even higher values of energy one expects
parton densities to become so large that the whole perturbative QCD
picture breaks down leading to a new regime of weak coupling but high
density non-perturbative QCD.  The goal in small-x physics is to
understand, experimentally and theoretically, the onset of BFKL
behaviour, its modification due to unitarity corrections as one
proceeds to higher energies and finally the nature of the transition
to the high density (saturation) regime at even higher energies.

From a theoretical point of view the simplest process to study is high
energy heavy quarkonium-heavy quarkonium scattering.  If the onium is
sufficiently heavy the radius of the onium is small enough to set the
single hard scale of the process.  Any understanding of BFKL dynamics
and beyond gained in studying heavy onium collisions should be
applicable to the more experimentally accessible single hard scale
processes mentioned above.  In the large $N_c$ limit of QCD a dipole
picture of high energy heavy onium-heavy onium scattering has been
developed which is equivalent to the usual BFKL description
\cite{Muel94a, MuPa94, Muel94b, Muel95, NiZZ94a}.  In this dipole
picture the light-cone wave function of a heavy onium is viewed as a
collection of colour dipoles.  The dipoles are made up of the quark
part of a gluon (or the heavy quark itself) and the antiquark part of
a different gluon (or the heavy antiquark).  High energy elastic
scattering in the BFKL approximation proceeds by two-gluon exchange
between a right-moving dipole in the right-moving onium and a
left-moving dipole in the left-moving onium.

Perhaps the main advantage of the dipole picture is that the equation
governing the number and distribution of dipoles in the light-cone
wave function of the onium is a branching equation which allows Monte
Carlo methods to be used.  In a recent paper \cite{Sala95b}, one of us
has studied high energy onium-onium scattering from energies where the
BFKL approximation is valid into the higher energy regime where
unitarity limits, for a fixed impact parameter of the scattering, are
reached.  In the centre of mass system unitarity corrections are
simply the independent scattering of two or more dipoles of the
right-moving onium with two or more dipoles of the left-moving onium.
Because of the large number of (small) dipoles in an onium wave
function unitarity corrections become important when the dipoles are
relatively dilute in the onium so that independent multiple
scatterings are the natural leading corrections to the BFKL
approximation.

The purpose of the present paper is to study two related questions,
the probability of producing an unusually large number of particles in
an onium-onium collision and the nature of gluon (dipole) evolution
when dipole densities become so dense that saturation effects due to
dipole-dipole interactions become important.  At a given energy dipole
densities in an onium wave function are largest not in a typical
configuration but rather in rare high multiplicity fluctuations.  Thus
attempts to measure saturation will naturally focus on the high
multiplicity tail of particle production.

One cannot directly study particle production in the dipole formalism.
In the dipole picture the $t=0$ light-cone wave function is
constructed.  This is sufficient to describe an elastic scattering,
but it is not sufficient to describe an inelastic event where one must
follow the time evolution of the wave function to a large positive
time when particles (or jets) are produced.  However, using the
Abramovskii, Gribov, Kancheli (AGK) cutting rules \cite{AbGK74} one
can relate particle production and forward elastic scattering
involving one or more dipole scatterings. While there has been no
detailed proof of the AGK rules in QCD it is likely that they are
correct and we assume that this is the case in the present paper.

We calculate $\sigma_k$ the cross section for $k$ cut pomerons when
$k$ takes values far above average.  The multiplicity of produced
particles, in the central unit of rapidity, should be proportional to
$k$. We begin by studying a toy model which obeys \emph{dipole}
evolution but where the absence of transverse dimensions makes
analytic solutions to the multiplicity distribution possible.  Our
main result for the toy model is given in eq.~(\ref{eq:toysigk}) where
$\sigma_k$ is seen to decrease exponentially in ${\sqrt{k}}$ at large
$k$, a tail much longer than that given by KNO scaling \cite{KoNO72}.
We have not been able to solve the large $k$ behaviour for the $k$ cut
pomeron contribution to a zero impact parameter collision in QCD,
$\df\sigma_k/\df^2\vr(\vr\simeq 0)$. However, Monte Carlo results
obtained using OEDIPUS (Onium Evolution, Dipole Interaction and
Perturbative Unitarisation Software) \cite{Sala96a} indicate a very
similar $k$-dependence to that found in the toy model with an
$e^{-A{\sqrt{k+n}}}$ dependence at large $k$. The same dependence, but
with different values of $A$ and $n$, also well describes onium-onium
collisions integrated over impact parameter.  The fact that the high
multiplicity tail extends so far encourages one to believe that high
parton density systems can be produced at reasonable accelerator
energies if the right signal is found to trigger on the interesting
physics associated with saturation.

In section 3, we turn to a study of saturation \cite{GrLR83},
beginning with defining measures of saturation.  Saturation effects
show up as a modification of the simple branching picture of dipoles
in constructing the square of the onium wave function.  Thus, a
non-independence of the dipoles in an onium wave function is a measure
of saturation.  However, it is not known how to estimate the
interaction of dipoles in an onium wave function directly. One expects
that the interaction of a slightly right-moving dipole with higher
rapidity right-moving dipoles in a right-moving onium should be
similar to that of a slightly left-moving dipole with the right-moving
onium.  This latter interaction is just the scattering amplitude for a
simple dipole with an onium state and is straightforward to calculate.
Thus one measure of saturation of a light-cone onium wave function is
the interaction probability with an opposite moving elementary dipole.
This measure of saturation is however time consuming to evaluate.  A
more efficient measure is the \emph{overlap} that a given dipole in
the onium wave function has with all the other dipoles in the wave
function.  Happily, it turns out that these two measures are strongly
correlated so that the simpler overlap measure can be used.

The toy model discussed earlier also furnishes a useful model of
saturation.  Saturation is expected to slow the growth of the
number of dipoles.  By comparing the cross section calculated in
frames where the two incoming particles share the incoming energy
differently, a consistency condition, requiring that the cross section
be frame independent, gives sufficient information to determine dipole
branching as a function of rapidity.  While at low energy the rate of
parton evolution grows exponentially in rapidity, as in BFKL
evolution, that rate reaches a constant (saturates) at high rapidity
as shown in eq.~(\ref{eq:toyrsn}).

The toy model can be used to study the intriguing, and still not well
understood, relationship between unitarity limits and saturation.  In
QCD the centre of mass onium-onium scattering, at a fixed impact
parameter, reaches the unitarity limit at energies well below that at
which saturation effects begin to appear.  However, if onium-onium
scattering is viewed in a frame where, say, the left-moving onium has
a small rapidity, the scattering looks exactly like the measure we have
earlier discussed for saturation since now the left-moving onium is
simply a single dipole.  Thus the multiple scatterings in the centre
of mass system must show up as saturation effects in a system where
one of the onia carries almost all the energy. This can be most
sharply seen in the toy model by considering a toy onium scattering on
a toy nucleus consisting of $N$ onia.  Here it is found that
saturation effects found earlier are exactly what is necessary to make
\emph{onium-nucleus} scattering the same in either the frame where the
onium is at rest or in the frame where the nucleus is at rest.

In QCD it is not possible to do such complete analytical calculations,
however, we have found a phenomenological modification of dipole
branching eq.~(\ref{eq:rlmdlstrn}) which makes onium-onium and
onium-nucleus scattering reasonably frame independent.  One
particularly sharp way to see the necessity of saturation effects and
their frame dependence is to compare the S-matrix, at zero impact
parameter, for onium-onium scattering in the centre of mass frame with
that obtained in a laboratory frame at very high rapidities.  At
rapidities large enough so that the S-matrix has become very small the
evaluation of the S-matrix neglecting saturation is determined by the
very low multiplicity fluctuations of the wave function.  We find, on
theoretical grounds and from numerical simulations, that $S\sim
e^{-c(Y-Y_0)^2}$. However, without saturation effects $c$ is a factor
of 2 larger in the lab frame than in the centre of mass.  Saturation
effects will slow the growth of dipoles in the laboratory frame so
that the typical wave function configuration agrees with the centre of
mass calculation coming from rare fluctuations.  This is exactly what
happens in the toy model as can be seen by comparing
eqs.~(\ref{eq:toysn}) and (\ref{eq:toysotyp}).

At large impact parameters, $r$, we find that the property of the
centre of mass frame, that saturation corrections set in much later
than unitarity effects, starts to break down. This is brought about by
several factors. At large $r$, the amplitude has a significant
contribution from asymmetric onium-onium configurations, for example
where one onium has a only central dense cluster of small dipoles (of
order of the onium size): for there to be an interaction the second
onium must have small dipoles at the relevant large impact parameter,
where they are likely to be much more dilute. This kind of scattering
looks similar to a lab frame collision, and hence the saturation and
unitarity corrections will be similar. A second point is that one
might expect a distribution of large dipoles which looks relatively
dilute, to have only small saturation corrections. However, the
sequence of evolution to produce this distribution will often have
passed through a stage with a dense collection of small dipoles, whose
evolution would be significantly altered by saturation effects,
reducing the production of large dipoles in the first place. Overall,
in the centre of mass frame, at large impact parameters, these effects
combine to give saturation corrections of the same order as the
unitarity corrections. Fortunately, the total scattering cross section
is not too strongly affected by these large impact parameters, and it
is still reasonable to neglect saturation effects in its calculation.

%----------------------------------------------------------------------
\section{Cuts}
\subsection{The AGK cutting rules}
It is most useful to start from the AGK cutting rules \cite{AbGK74} as
expressed in \cite{KoMu75}. The cross section, $\sigma_k$ involving
the exchange of exactly $k$ cut pomerons can be expressed as follows:

\begin{equation}
\sigma_k = \sum_{\nu = k}^{\infty} (-1)^{\nu - k} 2^\nu C^\nu_k
F^{(\nu)}, 
\label{eq:sigkf}
\end{equation}

\noindent where $F^{(\nu)}$ is the amplitude for the exchange of $\nu$
pomerons. In this notation, the total cross section $\sigma_t$ is

\begin{equation}
\sigma_t = 2 \sum_{\nu = 1}^\infty (-1)^{\nu - 1} F^{(\nu)}.
\label{eq:sigtotf}
\end{equation}

\noindent One knows from previous work \cite{Muel94b,Sala95b} that the
terms of the sum in eq.~(\ref{eq:sigtotf}) diverge as a factorial, and
therefore so will the terms in eq.~(\ref{eq:sigkf}). The reason for
this divergence is that rare configurations of the onia contribute
increasingly large amounts to the amplitudes for higher pomeron
exchange. The solution is to sum multiple pomeron amplitudes before
averaging over onium configurations. Applying this method here one
obtains the differential cross section for obtaining $k$ cut
pomerons. The $\nu$ pomeron amplitude for a fixed impact parameter
$\vr$ between the two onia (of sizes $\vb$ and $\vb'$, evolved to
rapidities $y$ and $Y-y$) is:

\begin{equation}
F^{(\nu)}(\vr, \vb, \vb', Y, y) = \frac{1}{\nu!} \sum_{\gamma, \gamma'}
        P_{\gamma}(\vr_0, \vb, y)
        P_{\gamma'}(\vr_0 + \vr, \vb',  Y-y)
	(f_{\gamma, \gamma'})^\nu,
\end{equation}

\noindent where $f_{\gamma, \gamma'}$ is the two gluon exchange
interaction amplitude between a pair of configurations $\gamma$ and
$\gamma'$ which occur with probabilities $P_{\gamma}(\vr_0, \vb, y)$
and $P_{\gamma'}(\vr_0 + \vr, \vb', Y-y)$ from onia situated at
$\vr_0$ and $\vr_0 + \vr$. Substituting this into eq.~(\ref{eq:sigkf})
gives

\begin{equation}
\frac{\df^2 \sigma_k}{\df^2 \vr} = 
\frac{1}{k!} \sum_{\gamma, \gamma'}
	P_{\gamma}(\vr_0, \vb, y)
        P_{\gamma'}(\vr_0 + \vr, \vb',  Y-y)
	(2f_{\gamma, \gamma'})^k \exp(-2f_{\gamma, \gamma'}).
\label{eq:sigkp}
\end{equation}

\noindent The cross section for obtaining no cuts is the difference
between the unitarised cross section and $\sum_{k=1}^{\infty}
\sigma_k$, which comes out as:

\begin{equation}
\frac{\df^2 \sigma_0}{\df^2 \vr} = 
	\sum_{\gamma, \gamma'}
	P_{\gamma}(\vr_0, \vb, y)
        P_{\gamma'}(\vr_0 + \vr, \vb',  Y-y)
	\left(1 - \e^{-f_{\gamma,\gamma'}}\right)^2.
\end{equation}

\noindent In the work that follows, $y$ will be set to $Y/2$, so that
we will be examining the number of cuts at a central rapidity.

\subsection{Toy model calculation}
As was the case for the unitarised amplitude \cite{Muel94b}, the
evaluation of eq.~(\ref{eq:sigkp}) is very difficult to do
analytically, because it requires a detailed understanding of the
distribution of dipole configurations for an onium. However some of
the features of multiple cut cross sections can be extracted by
examining a ``toy'' model which has no transverse dimensions
\cite{Muel94b}.

The main feature of the toy model is that the probability of obtaining
a large number $n$ of dipoles has an exponential
distribution, $P_n \simeq e^{-n/\mu} / \mu$, where $\mu$ is the mean
number of dipoles (proportional to $e^{\apm y}$ for an onium evolved to
a rapidity $y$). One can show that in the case with transverse
dimensions, the probability of obtaining a large density of dipoles
also has a distribution which is an exponential in the density
\cite{Sala95b}. It is this similarity which leads to the
correspondence between the toy model and the case with two transverse
dimensions. In the toy model, the cross section for an interaction
with $k$ cut pomerons is

\begin{equation}
\sigma_k = \frac{1}{k!} \sum_{m,n} (2f\alps^2 nm)^k
	\exp(-2f\alps^2 nm) P_m P_n,
\end{equation}

\noindent where the interaction amplitude between two dipoles is
defined to be $f\alps^2$ (equivalent to $f_{\gamma, \gamma'}$
above). There are two limits which can be taken. Let $F^{(1)}$ be the
amplitude for exchange of one pomeron. If $kF^{(1)} = k f \alps^2
\mu^2 \ll 1$, then unitarisation effects (the exponential factor) are 
unimportant, and the evaluation of the cuts cross sections is very
similar to the evaluation of the multiple pomeron amplitudes
\cite{Muel94b}: 

\begin{equation}
\sigma_k \simeq k! (2F^{(1)})^k \e^{-2k^2F^{(1)}}.
\end{equation}

\noindent The opposite limit is when unitarisation effects are
important, $kF^{(1)} = k f \alps^2 \mu^2 \gg 1$ (and also $k/F^{(1)}
\gg 1$). Approximating the sums by integrals, and changing variables
to $n = \sqrt{u} \e^\xi$ and $m = \sqrt{u} \e^{-\xi}$ gives

\begin{equation}
\sigma_k \simeq \frac{1}{\mu^2 k!} (2f\alps^2)^k
	\int \df u \df \xi u^k \e^{-2f\alps^2u}
	\e^{-2(\cosh \xi) \sqrt{u}/\mu}.
\end{equation}

\noindent Performing the $\xi$ integration using the saddle point
approximation, it is clear that the integral is dominated by a region
where $m$ and $n$ are similar. Performing the second integration gives

\begin{equation}
\sigma_k \simeq \frac{1}{2F^{(1)}}
	\left(\frac{2F^{(1)}\pi^2}{k}\right)^{1/4}
	\exp\left(-\sqrt{\frac{2k}{F^{(1)}}}\right).
\label{eq:toysigk}
\end{equation}

\noindent The unusual feature of this result is the square root
dependence on $k$ in the exponential. Given that the number of cuts is
related to the final state multiplicity, one might have expected the
usual KNO type \cite{KoNO72} exponential distribution. The origin of
the $\sqrt{k}$ behaviour is that, in the region of configuration space
contributing to the $k$-cut amplitude, the factor $(2f\alps^2 nm)^k
\exp(-2f\alps^2 nm)$ is the most rapidly varying and fixes the dominant
values of $n,m \propto \sqrt{k}$. The probabilities associated with
the relevant configurations just fix the normalisation, with
$P(\sqrt{k}) \sim \e^{-\sqrt{k}}$.

\subsection{Monte Carlo results for differential cut cross sections}
One is most likely to find a similarity between the toy model results
and the realistic situation, by examining the differential cut cross
sections, which eliminates two of the transverse degrees of
freedom. In \cite{Sala95b} correspondence between the toy model and
Monte Carlo results for zero impact parameter has already been
established for the multiple pomeron exchange amplitudes. The main
difference was that where the quantity $\nu$ occurred in an expression
relating to $\nu$-pomeron exchange, it had to be replaced with $(\nu +
n)$ where $n$ was a constant which originated from the details of the
dynamics in the transverse dimensions. One might expect a similar
effect when looking at cut cross sections, and one can examine the
differential cross section for small impact parameters for behaviour
of the form

\begin{equation}
\frac{\df^2\sigma_k}{\df^2 \vr} (r \simeq 0) \simeq \exp[-A(k+n)^\gamma
+ B] 
\label{eq:sqrtfit}
\end{equation}

\noindent where $A$, $B$, $n$ and $\gamma$ are constants to be
determined. The first stage is to ensure that one really does have
$\gamma = 1/2$. By defining $l_k = - \ln\df^2\sigma_k/\df^2\vr$ and
plotting 

\begin{equation}
\Gamma_k = \frac{l_{k+1} - l_{k-1}}{2l_k - l_{k-1} - l_{k+1}}
= \frac{2(k + n)}{1 - \gamma} + O(1/(k+n))
\end{equation}

\begin{figure}
\begin{center}
\epsfig{file=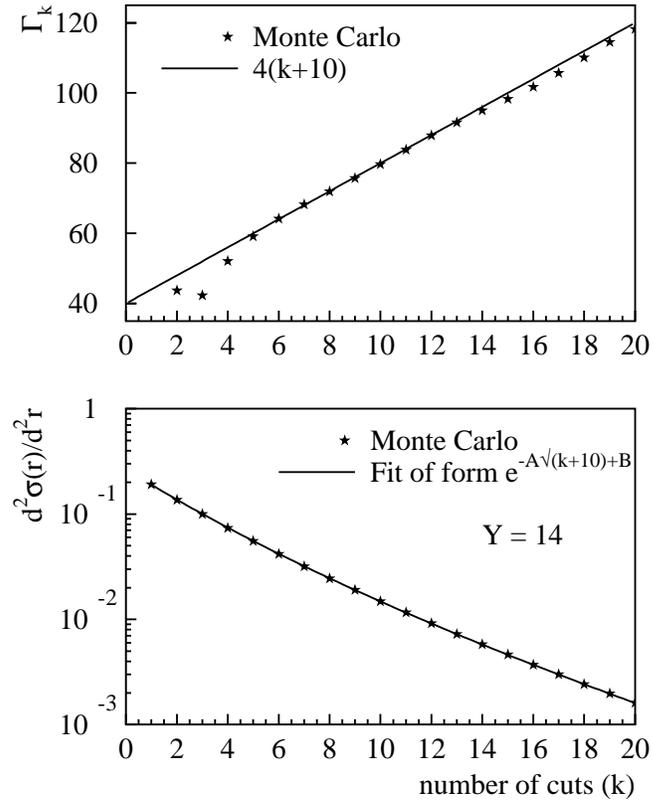, width = 0.6\textwidth}
\caption{The upper plot shows $\Gamma_k$ for fixed impact
parameter: a test that the cut cross sections show the behaviour
corresponding to eq.~(\ref{eq:sqrtfit}) (actually averaged over a
range of $r = 0 \to 0.6b$); the lower plot shows the multiple cut cross
sections together with a fit of the form eq.~(\ref{eq:sqrtfit}) (for
the range $r = 0.2b\to 0.3b$).}
\label{fig:rbc_v_sqrt}
\end{center}
\end{figure}

\noindent (essentially the ratio of first and second derivatives) one
can eliminate $A$ and $B$. Figure~\ref{fig:rbc_v_sqrt} shows this
quantity as a function of $k$, determined using the OEDIPUS Monte
Carlo simulation \cite{Sala96a}. For $k$ sufficiently large, the
linear behaviour is very clear, and the slope corresponds to $\gamma =
1/2$. The value of the intercept with the $k$ axis yields $n$, which
has a value of $10$. This is somewhat larger than the corresponding
value found for multiple pomeron exchange ($\nu \to \nu + n$, with
$n\sim 3$). The difference can be explained by arguing that the
multiple pomeron and multiple cut cross sections are sensitive to the
dipole distributions (and the manner in which they differ from an
exponential) in different ways.

The other property to examine is the variation of $A$ with $Y$. One
clearly cannot expect $A$ to be exactly $\sqrt{2/F^{(1)}(0)}$, however
it might be reasonable for it to be proportional to
$\sqrt{2/F^{(1)}(0)}$. Table~\ref{tbl:AvF} gives $A$ for two values of
$Y$, indicating that to within the accuracy of the Monte Carlo results,
this proportionality does hold.

\begin{table}[h]
\begin{center}
\begin{tabular}{|c|c|c|c|c|} \hline
 $Y$ & $n$ & $A$ & $F^{(1)}(0)$ & $A/\sqrt{2/F^{(1)}(0)}$ \\ \hline
 10  & 7   & 4.45 & 0.436 & 2.08  \\ \hline
 14  & 10  & 2.21 & 1.62  & 1.99  \\ \hline
\end{tabular}
\end{center}
\caption{Parameterisations of cut distributions at central impact
parameters.}
\label{tbl:AvF}
\end{table}

\noindent The dependence of the distributions on $n$ is relatively
weak which is why it is not determined more
accurately. Table~\ref{tbl:AvF} therefore confirms that for the zero
impact parameter cross section the distribution scales as suggested by
the toy model.

\subsection{Monte Carlo results for the integrated cut cross sections}
\begin{figure}
\begin{center}
\epsfig{file=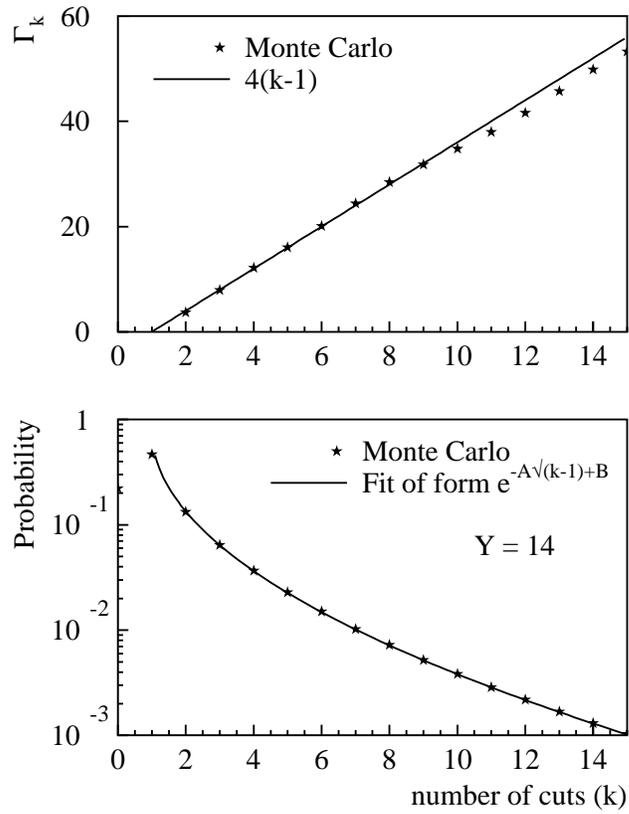, width = 0.6\textwidth}
\caption{The upper plot shows $\Gamma_k$ for the integrated cross
sections; the lower plot shows the multiple cut cross
sections together with a fit of the form eq.~(\ref{eq:sqrtfit}).}
\label{fig:tc_v_sqrt}
\end{center}
\end{figure}

One can argue that events with $k$ cut pomerons will have a
multiplicity proportional to $k$. In an experimental situation, one
might be able to determine the cross sections for various ranges of
multiplicities, but this will correspond to cut cross sections
integrated over impact parameter, for which there will not necessarily
be any correspondence with the toy model. However
figure~\ref{fig:tc_v_sqrt} shows that a fit of the form
eq.~(\ref{eq:sqrtfit})

\begin{equation}
\sigma_k \simeq \exp[-A(k+n)^\gamma + B]
\end{equation}

\noindent works remarkably well. The value of $n$ turns out to be
considerably smaller than for eq.~(\ref{eq:sqrtfit}): this is
associated with the fact that at larger values of $r$ the value of $n$
decreases.

\begin{figure}
\begin{center}
\epsfig{file=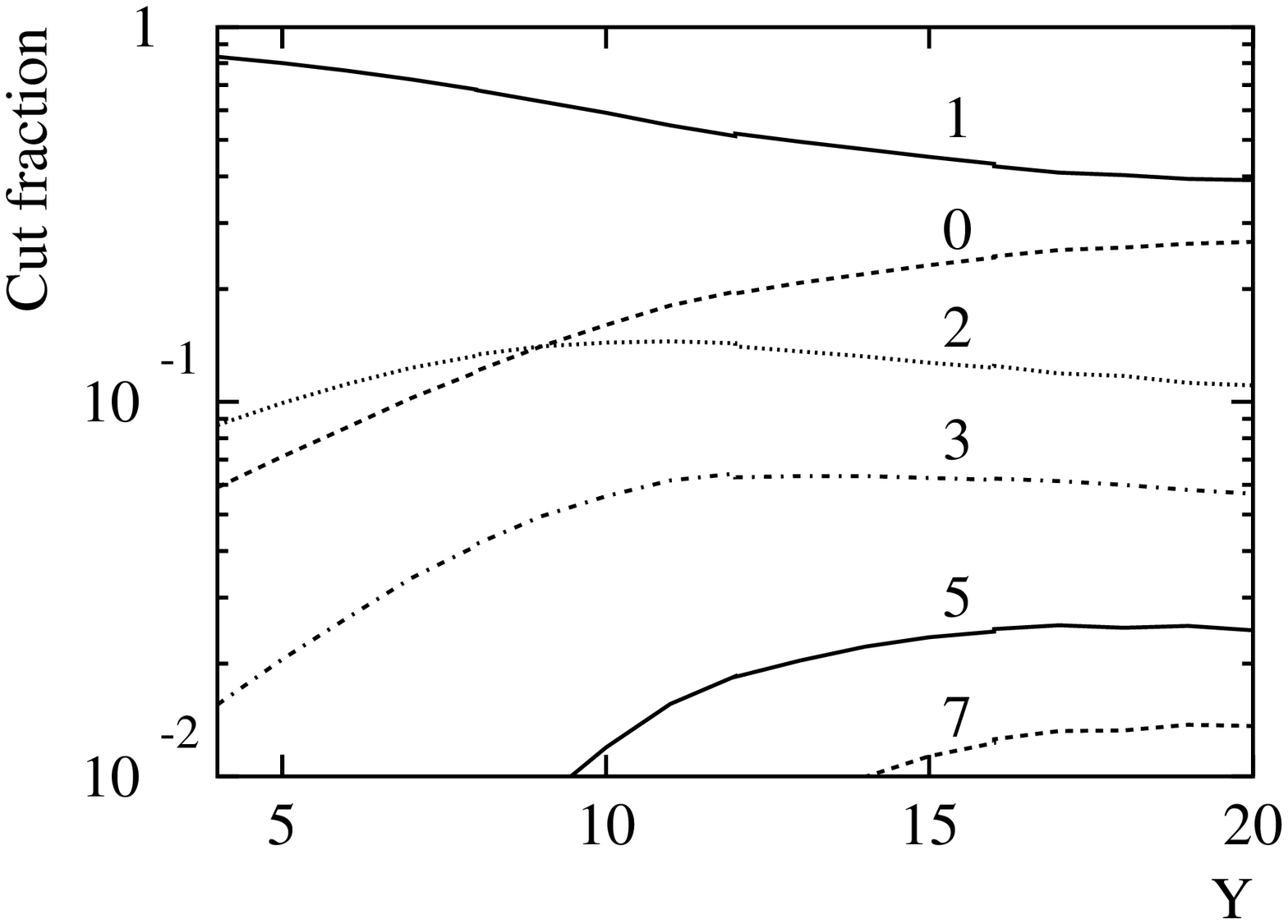, width = 0.6\textwidth}
\caption{The fraction of the total cross section coming from different
numbers of cuts (indicated by the labels next to the curves) as a
function of rapidity.}
\label{fig:allcuts_frac}
\end{center}
\end{figure}

It is not clear why the $\exp(-A\sqrt{k+n})$ behaviour also applies to
the integrated cross section. The results do however suggest that the
variation of the coefficient $A$ with the rapidity is not as simple as
in the toy model (or in the fixed impact parameter case).  The
numerical results for the variation with rapidity of the fraction of
the cross section coming from different numbers of cuts is shown in
figure~\ref{fig:allcuts_frac}.

\subsection{Spatial distribution of cut cross sections}
\begin{figure}
\begin{center}
\epsfig{file=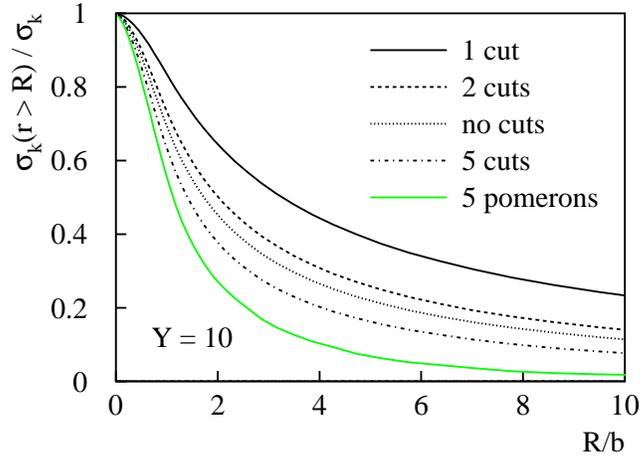, width = 0.6\textwidth}
\caption{The fraction of the multiple cut cross section which comes
from impact parameters larger than $R$ for a variety of numbers of
cuts. At this rapidity, the mean number of cuts is $\sim 1.3$}
\label{fig:cmltv_cts}
\end{center}
\end{figure}

Just as high numbers of pomeron exchange tend to be dominated by small
impact parameters, one can expect events with large numbers of cuts
also to be relatively central, because they are dominated by
dense configurations, which arise most commonly at small impact
parameters. This is illustrated in figure~\ref{fig:cmltv_cts}, which
shows $\sigma_k(r> R) / \sigma_k$, the fraction of the $k$-cut cross
section which comes from impact parameters larger than $R$, for
various numbers of cuts, as a function of $R$.

One sees for example that about half the one-cut cross section is
coming from a region $r < 3.2b$, whereas the first half of the 5-cut
cross section comes from $r < 1.4b$. There is of course a limit to how
central the many cut cross section can become, since any cross section
is bound to be spread out over a region at least the size of the
onia. The no cut cross section is also quite central: like the many
cut cross sections, it is particularly sensitive to dense
configurations. Note that all these cross sections have significant
tails at large $r$. This is to be contrasted with the 5-pomeron curve,
where the tail dies off quickly: the difference arises because of the
unitarity corrections that are associated with multi-cut cross
sections but not multi-pomeron contributions, and which limit the
relative contributions from very dense configurations.

%----------------------------------------------------------------------
\section{Saturation}
In the calculations performed to determine the unitarity corrections
\cite{Muel94b, Sala95b}, it has been assumed that saturation of the
wave function can be neglected. The justification for doing this
is that in a situation where both wave functions are evolved to $Y/2$,
unitarisation corrections will set in when $\alps^2 e^{\apm Y} \sim
1$, while saturation corrections for a particular dipole will be of
the order $\alps^2 e^{\apm Y/2}$. Two problems can arise with this
argument. The first relates to the question of whether $\alps$ is
small enough. In the calculations presented here, $\alps \sim 0.2$, so
when unitarisation is important one might expect $\alps^2 e^{\apm Y/2}
\sim 0.2$. Since the conditions for saturation can only be guessed to
within a factor of order 1, one is potentially close to the onset of
saturation corrections as well. Secondly, the argument is based on the
principle that the mean dipole density will be a good gauge of the
saturation corrections. However in the case of unitarisation
corrections, it has been shown \cite{Sala95b} that the use of mean
interactions can be very misleading. In this section we will perform a
more sophisticated analysis of the importance of saturation
corrections, relative to unitarity corrections. We will also present
work on understanding wavefunction evolution once saturation becomes a
large effect.

\subsection{Measures of saturation}
It is first necessary to develop a measure of saturation. Two methods
will be used here. Firstly one can probe a wave function with an
(unevolved) onium. The two gluon exchange amplitude between the wave
function and the onium indicates the local density dipoles of size
similar to the onium and might be similar to the interaction that a
dipole in the wave function would have with its neighbours. Since the
dipole-dipole interaction has spikes wherever two sources overlap, one
should average the interaction over a region of roughly the onium
size. The choice of the size of the region over which to average is
one of the main uncertainties with this method. It also has the
disadvantage of being relatively slow, because one needs to determine
dipole-dipole interactions at many points to determine the saturation
over the whole extent of the wave function and if one wants the
saturation associated with dipoles of different sizes, one must use
probes of different sizes.

The second method is more arbitrary, but much faster to compute. Each
dipole has an ``overlap'' associated with it.  For each pair of
dipoles within the onium wave function which are separated by less
than a certain distance (which will be a function of the dipole sizes)
the quantity $\alps^2 c_<^2 / c_>^2$ is added to the overlap for each
of the two dipoles (where the smaller dipole has size $c_<$ and the
larger, size $c_>$). This function is loosely based on the fact that
the total two-gluon interaction amplitude between a pair of dipoles
(moving fast relative to each other)  has the form $\alps^2 c_<^2 /
c_>^2(1 + \ln c_>/c_<)/2$ and that it is dominated by a region where
the two dipoles are close in impact parameter. This method has the
advantage that one obtains a value of the saturation for each dipole.

\begin{figure}
\begin{center}
\epsfig{file=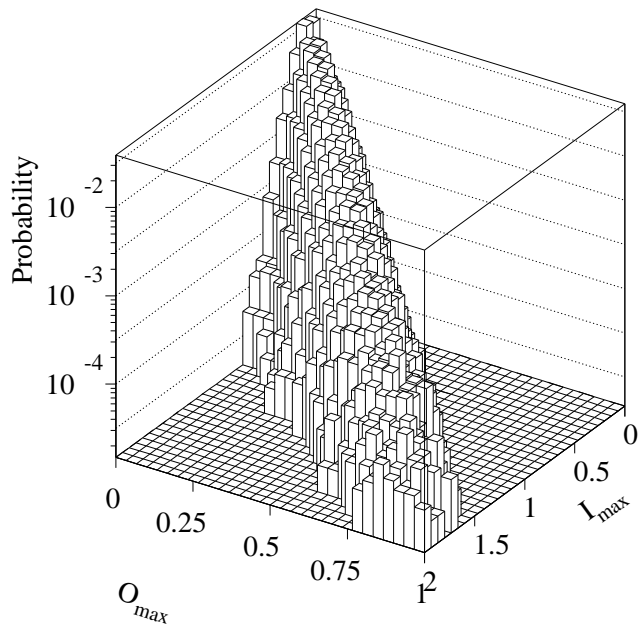, width = 0.6\textwidth}
\caption{The probability of a given onium configuration lying in a
certain $I_{max}$, $O_{max}$ bin, for $y=10$.}
\label{fig:op_histo}
\end{center}
\end{figure}

Comparing the two methods one finds that there is a good (linear)
correlation between them, as can be seen in figure~\ref{fig:op_histo}
which shows the frequency distribution for configurations with various
values of $I_{max}$ (the maximum interaction with the probe) and
$O_{max}$ (the maximum overlap). One can carry out other comparisons,
in particular, considering the location in impact parameter of the
maximum saturation, and one again finds that the two methods give very
similar results. In addition, if one examines other values of
rapidity, one finds that the factor relating $I_{max}$ and $O_{max}$
stays approximately the same.

\subsection{Distribution of $I_{max}$}
The degree of saturation is obviously just related to the density of
dipoles. Therefore the distribution of $I_{max}$ should be related to
the probability of obtaining large densities, which has been shown
to be exponential \cite{Sala95b}, the slope of the exponential being
proportional to the mean density of dipoles. Instead of searching the
whole relevant region of impact parameter for the location of
$I_{max}$, one finds the location of the dipole corresponding to
$O_{max}$ and bases the search for $I_{max}$ around that point. The
probe size is always the same as the onium size.

One finds, as expected, that $I_{max}$ also shows an exponential
behaviour, with a slope varying in proportion to the local density of
dipoles. One also finds that the maximum saturation is usually close
to the original onium which is also unsurprising since that is where
the evolution has the longest rapidity range to produce a large number
of dipoles.

While these properties are of general interest, one of the main reason
for studying saturation is to see whether it is likely to affect the
onium-onium cross section. So one needs to know the typical values of
$I_{max}$ and also the correlation between $I_{max}$ for a pair of
onium configurations and the interaction cross section for that pair
of configurations.

\begin{figure}
\begin{center}
\epsfig{file=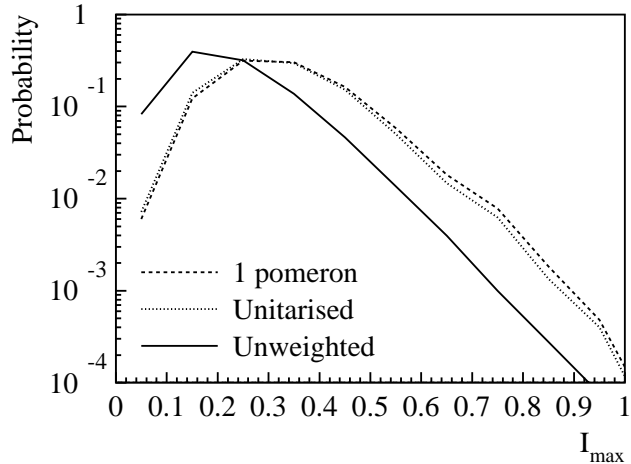, width = 0.6\textwidth}
\caption{The probability distribution of the larger $I_{max}$ from a
pair of onia (`unweighted' curve), and the fractions of cross sections
coming from configuration pairs with particular values of $I_{max}$
(`1-pomeron' and `unitarised' curves); $Y=12$.}
\label{fig:sig_nw}
\end{center}
\end{figure}

The most appropriate value of rapidity to examine is $Y=12$, since
this is the rapidity for which the mean 1-pomeron amplitude at $r=0$,
$F^{(1)}(r=0)$, is equal to 1 (the unitarity bound). Consider first
the `unweighted' curve of figure~\ref{fig:sig_nw}: this corresponds to
the probability distribution of $I_{max}$ in onium-onium collisions
with a total rapidity $Y=12$. The rapidity for each onium is $y = Y/2
= 6$. $I_{max}$ is now defined to be the larger of the $I_{max}$
values for the two onium configurations. One sees that typical values
of saturation are $0.1 \to 0.3$. The exponential probability
distribution for large saturation is relatively clear. More relevant
though are the distributions weighted with the onium-onium cross
section for each configuration pair: the weighted distributions
correspond to the fraction of the cross section coming from
configuration pairs with a certain $I_{max}$, and one sees that the
typical saturation has increased to be in the range $0.2 \to
0.45$. There are two main reasons why configuration pairs with
larger cross sections are associated with larger saturation. Firstly
if one of the onia has a high density of dipoles in some region, then
the interaction of that region with a second onium will be enhanced. A
second possibility, which we will consider in more detail later, is
that configurations which lead to a large interaction (e.g. those with
large dipoles) are most likely to be produced through branching
sequences which pass through a high density stage.

There remains the problem of assessing the effect that this degree of
saturation would have. One can assume that saturation will start to
become important when $I_{max}\sim 1$, but this is a very approximate
condition, being uncertain to within at least a factor of two. This
factor of two is critical: if the condition for saturation is $I_{max}
\sim 2$ then there will be no problem with saturation at $Y=12$, on
the other hand if it is $I_{max} \sim 1/2$ then saturation could
significantly alter the results at this rapidity. If saturation is
important then one wants to know how it will affect the evolution: for
example the saturation might be large in one region of impact
parameter, but the bulk of the cross section might come from different
regions, not affected by saturation. The next sections address
these and other issues.

\subsection{Saturation in the toy model}
One approach to the problem is again from the toy model, using
the condition that the scattering amplitude must be independent of the
frame in which calculation is performed. Equivalently, it shouldn't
matter which of the two onia one chooses to evolve. Starting with a
state containing $n$ dipoles, the rate of branching into a state
containing $n+1$ dipoles is defined to be $R_n$, while the interaction
between two states with $n$ and $m$ dipoles respectively is $F(n,m)$.
If the evolution is partitioned such that for an increase $\delta Y$
in total rapidity, the onium with $n$ dipoles evolves by $\xi \delta
Y$ and the one with $m$ dipoles evolves by $(1-\xi) \delta Y$, then
the evolution of the amplitude $F$ will be

\begin{equation}
\frac{\df F}{\df Y} = \xi R_n \left[F(n+1,m) - F(n,m) \right] + 
		    (1-\xi) R_m \left[F(n,m+1) - F(n,m) \right]
\label{eq:toyinv1p}
\end{equation}

\noindent In the case of one pomeron exchange, one has $R_n = \apm n$
and $F(n,m) = f \alps^2 nm$ (where $f$ sets the strength of the
interaction), giving 

\begin{equation}
\frac{\df F^{(1)}}{\df Y} = [\xi + (1- \xi)]  \apm f \alps^2 nm 
\end{equation}

\noindent which is clearly independent of $\xi$.

If for $F(n,m)$ one uses the form which includes unitarity
corrections, $F(n,m) = [1 - \exp(-f\alps^2 nm)]$, one obtains the
following expression for the evolution of the interaction:

\begin{equation}
\frac{\df F^s}{\df Y} =
 \left[\xi R^s_n \left(1 - \e^{-f\alps^2m}\right) + 
	(1-\xi) R^s_m \left(1 - \e^{-f\alps^2n}\right) \right]
	e^{-\alps^2 mn}
\label{eq:toyinv}
\end{equation}

\noindent where the superscript $s$ indicates that
saturation is being taken into account. The condition that the growth
of the cross section be independent of $\xi$, as well as one's
knowledge of the branching rate for small $n$ (where saturation should
be unimportant) leads to the following expression for the rate of
branching:

\begin{equation}
R^s_n = \frac{\apm}{f\alps^2}\left(1 - \e^{-f\alps^2n}\right).
\label{eq:toyrsn}
\end{equation}

\noindent Note that though the interaction with a second onium was
used in the determination of this branching rate, the result is
independent of the state of the second onium.

To see what effect saturation will have on the final probability
distribution of dipoles, one can use the following arguments. First
take the case without saturation, in the continuum limit for $n$. The
probability distribution satisfies

\begin{equation}
\frac{\df P}{\df y} = - \frac{\df R_n P}{\df n} =
 -\apm \frac{\df nP}{\df n}.
\end{equation}

\noindent Changing variables to $Q = nP$, $\zeta = \apm y$ and $\nu =
\ln n$, one has  

\begin{equation}
\frac{\df Q}{\df \zeta} = - \frac{\df Q}{\df \nu},
\label{eq:qnu}
\end{equation}

\noindent whose solution (from our knowledge that $P = \exp[-\apm y -
n\exp(-\apm y)]$) is

\begin{equation}
Q = \exp\left( \nu - \zeta - \e^{\nu - \zeta}\right).
\end{equation}

\noindent In the case with saturation one has 

\begin{equation}
\frac{\df P^s}{\df y} = -\frac{\df R_n^s P^s}{\df n} =
 -\frac{\apm}{f\alps^2} \frac{\df (1 - \e^{-f\alps^2n})P^s}{\df n}.
\end{equation}

\noindent By an appropriate change of variables, $Q^s = (1 -
\e^{-f\alps^2n})P^s/(f\alps^2)$, $\zeta = \apm y$ and $\nu^s = \ln
[(\e^{f\alps^2n} - 1)/(f\alps^2)]$, one can obtain an equation
identical to eq.~(\ref{eq:qnu})

\begin{equation}
\frac{\df Q^s}{\df \zeta} = - \frac{\df Q^s}{\df \nu^s}.
\end{equation}

\noindent Because, in the limit of small $n$ and $\zeta$, $Q^s \simeq
Q$ and $\nu^s \simeq \nu$, the equation for $Q^s$ will have the same
initial conditions as that for $Q$ and therefore their solutions will
be the same. Translating this back to the variables $P^s$ and $n$, one
has

\begin{equation}
P^s(n,y) = \exp\left[ f\alps^2n - \apm y - \frac{1}{f\alps^2}
\left(\e^{f\alps^2 n} -1 \right) \e^{-\apm y} \right].
\label{eq:pssln}
\end{equation}

\noindent For $\e^{\apm y}, n \ll 1/(f\alps^2)$ this approximates to
the solution without saturation. The main feature of this solution
though is that at large $y$, instead of there being an exponential
distribution in $n$ with mean $\exp(\apm y)$, one now has a solution
which has a maximum at 

\begin{equation} 
n = \frac{1}{f\alps^2} \left[ \apm y + \ln f\alps^2 \right],
\label{eq:strnmaxn}
\end{equation}

\noindent and which dies off quickly beyond that.\footnote{Note
that beyond $n = \apm y/f\alps^2$ the solution eq.~(\ref{eq:pssln})
breaks down because the continuum limit used in its derivation is no
longer valid ($P^s$ varies too rapidly over a range $\delta n \sim
1$). Numerical solution shows though that there is still a rapid fall
off beyond the maximum, having roughly a Poisson distribution, since
the branching rate $n \to (n+1)$ becomes independent of $n$.}

One effect of this modified probability distribution is to increase
the value of the $S$-matrix at very large rapidities. If one
calculates the $S$-matrix in a centre of mass frame (rapidity $Y$
divided equally between the two onia) neglecting saturation one
obtains \cite{Muel94b}:

\begin{equation}
S(Y) \simeq \frac{-1}{f\alps^2\e^{\apm Y}} 
 \ln \left[f\alps^2\left(1 + \frac{1}{f\alps^2 \e^{\apm
Y/2}}\right)^2 \right].
\label{eq:toy_cm_ns}
\end{equation}

\noindent Note that this is much larger than the value of the
$S$-matrix that one obtains from typical configurations (namely
$\exp(-f\alps^2e^{\apm Y})$). This difference arises because the
$S$-matrix is dominated by a small fraction of configurations which
have few dipoles.

Calculating the $S$-matrix including saturation (done in the lab
frame, to reduce sensitivity to the use of the continuum limit, though
the result is of course independent of frame), one obtains

\begin{equation}
S(Y) \simeq \frac{1}{f\alps^2\e^{\apm Y}}
\left[ \ln f\alps^2 \e^{\apm Y} \right].
\label{eq:toy_cm_st}
\end{equation}

\noindent When $f\alps^2 \e^{\apm Y/2} \ll 1$,
eqs.~(\ref{eq:toy_cm_ns}) and (\ref{eq:toy_cm_st}) give the same
result, consistent with the idea that saturation should not be
important in the centre of mass frame for that energy range. However
for $f\alps^2 \e^{\apm Y/2} \gg 1$, eqs.~(\ref{eq:toy_cm_ns}) and
(\ref{eq:toy_cm_st}) do differ (the saturated case has a logarithmic
dependence on energy, absent from the calculation without saturation),
because saturation then matters in the evolution of the onium
wave functions, even in the centre of mass frame.

While this approach works well for the toy model, in the real case, an
analogous method will turn out to be too complex to be
useful. Nevertheless one can also make use of the effect of a change
of frame on the average (unitarised) amplitude. The simplest pair of
frames to examine would be centre of mass and lab frames for
onium-onium scattering. However one finds that, except for logarithmic
factors, the $S$-matrix behaves in a very similar way in the
two. Consider instead collisions between a toy onium and a toy
``nucleus'' consisting of $N$ onia. Two frames will be used: a lab
frame $L_O$ where the onium is at rest, and a lab frame $L_N$ where
the nucleus is at rest. At high energies (neglecting saturation), in
frame $L_N$, with the onium being evolved, the matrix is approximately

\begin{equation}
S_N \simeq \frac{1}{N f \alps^2 \e^{\apm Y}}.
\label{eq:toysn}
\end{equation}

\noindent Except for logarithmic terms (and the factor $N$) this has
the same leading energy behaviour as the centre of mass onium-onium
$S$-matrix, and is similarly dominated by those rare configurations of
the onium which have few dipoles (the common configurations have such
a large number of dipoles that their contribution to the $S$-matrix
can be neglected).

In the $L_O$ frame, it is the nucleus which it is evolved. Assuming
for the sake of the argument that the constituent onia evolve
independently (consistent with our neglect of saturation),
configurations with few dipoles will be much rarer, because they
require that none of the $N$ constituent onia should have evolved
much. Accordingly, when one calculates the $S$-matrix, one finds

\begin{equation}
S_O \simeq \frac{1}{\left( f\alps^2 \e^{\apm Y} \right)^N},
\end{equation}

\noindent which is much smaller than the result in the $L_N$
frame. One way of restoring the correspondence is by requiring
that saturation alter the probability distribution of the number
of dipoles so that the value of the $S$-matrix for a \emph{typical}
configuration of the nucleus is now 

\begin{equation}
S_O^{\mathrm{typical}} \sim \frac{1}{N f \alps^2 \e^{\apm Y}}.
\label{eq:toysotyp}
\end{equation}

\noindent Equivalently the typical number of dipoles must be of the
order of $[\apm Y + \ln Nf\alps^2]/ (f\alps^2)$. Note that this is
just the result of eq.~(\ref{eq:strnmaxn}) (except for the $\ln N$
term which is a result of the use of nuclei). The advantage of
examining the effect of frame changes on the average $S$-matrix is
that one can transfer the arguments fairly straightforwardly to the
case of scattering with transverse dimensions.

\subsection{Saturation in the case with transverse dimensions}
The real case, including transverse dimensions, is very much more
complicated, to a large extent because the dipole picture breaks down
at high dipole densities and one needs to take into account
quadrupoles, hexapoles, and other more complicated colour structures
\cite{Muel95,KoMW96}. This section will deal with various approaches that one
can nevertheless try in order to deduce more information about the
effects of saturation in the real case, still using the requirement of
frame independence.

It is simple to write down the equation which must be satisfied for
frame independence to hold. This will be the equivalent (with
transverse dimensions) of the independence of eqs.~(\ref{eq:toyinv1p})
and (\ref{eq:toyinv}) on $\xi$. Let $\gamma$ be a particular gluon
configuration and $\{\gamma_\vb\}$ be the set of all colour
configurations with an extra gluon added at transverse position $\vb$,
with $\gamma_\vb$ being any particular one of these. In the dipole
approximation (large $N_c$), a particular $\gamma_\vb$ corresponds to
the gluon having originated from a particular dipole. Define also
$F_{\gamma,\gamma'}(\vr)$ to be the interaction between configurations
$\gamma$ and $\gamma'$ (moving in opposite directions) when they have
a relative separation in impact parameter of $\vr$. Then for the
scattering to be frame independent (or equivalently for it not to
matter which onium is evolved) the rate of evolution $R(\gamma \to
\gamma_\vb)$ must satisfy

\begin{equation}
\sum_{\{ \gamma_{\vb}\}} \int \df^2\vb R(\gamma \to \gamma_\vb)
	\left(F_{\gamma_\vb, \gamma'} - F_{\gamma, \gamma'} \right)
 = 
\sum_{\{\gamma_\vb'\}} \int \df^2\vb R(\gamma' \to \gamma_\vb')
	\left(F_{\gamma, \gamma_\vb'} - F_{\gamma, \gamma'} \right)
\label{eq:realinv}
\end{equation}

\noindent In the case of of the one-pomeron approximation for the
scattering amplitude $F$, and the normal dipole evolution kernel for
$R$, if one can show that eq.~(\ref{eq:realinv}) holds for a single
dipole in each onium, then because dipoles in each onium interact (and
evolve) independently of one another eq.~(\ref{eq:realinv})
automatically holds for any collection of dipoles in each onium. The
fact that eq.~(\ref{eq:realinv}) holds for the 1-pomeron case is a
result of the conformal invariance of the dipole-dipole interaction
and of the dipole evolution kernel. The details are given in the
appendix.

The effect of including multi-pomeron interactions is to connect the
interaction of different dipoles within the same onium, and as a
result dipoles no longer necessarily evolve independently of one
another: this is the breakdown of the dipole picture of evolution
which necessitates that one take into account quadrupoles and more
complicated terms. Unfortunately it has not so far been possible to
solve eq.~(\ref{eq:realinv}) to obtain detailed information on the
effects of saturation.

One can nevertheless use frame independence to deduce some limited
information. We will first examine the effect of saturation on central
impact parameters. The argument will be along similar lines to the
toy-model approach discussed at the end of the last section, except
that now it will just be sufficient to consider simple onium-onium
scattering in lab ($L$) and centre of mass ($CM$) frames. It has been
shown in \cite{Sala95b} that there is a close correspondence between
toy model and real results for the probability distribution of the
number of dipoles when this number is high compared to the
mean. However the arguments used here depend on the probabilities of
having unusually low dipole densities, where the toy model is a poor
approximation to the real case. In a lab frame one finds that the
$S$-matrix is approximately $S_L \propto \e^{-c(Y-Y_0)^2}$ instead of
$S_L \propto \e^{-c (Y-Y_0)}$ in the toy model. Qualitatively, the
difference can be understood as coming from the extra dynamics
associated with the diffusion in dipole sizes. In the toy model, the
probability of having no evolution is just the exponentially
suppressed probability of a single dipole not branching over a
rapidity $Y$. In the real case the probability ($P_\rho(Y)$) of a
dipole of size $b$ not producing any new dipoles larger than a small
size $\rho$ is

\begin{equation}
P_\rho(Y) = \exp \left(-\frac{2\alps N_C Y}{\pi} \ln \frac{b}{\rho}
\right).
\end{equation}

\noindent The extra element which arises is in the choice of $\rho$,
since in the rapidity interval $Y$ one must also require that none of
the dipoles produced below size $\rho$ branch back up into ones of
size $b$. Forcing the probability of producing an extra large dipole
from the small dipoles to be small (approximately $e^{\apm Y}
\rho^2/b^2 \ll 1$), one obtains the condition $\ln b/\rho \propto Y$
which leads to the observed dependence of the $S$-matrix on $Y^2$
(since the $S$-matrix is dominated by the configurations with small
interactions and therefore few dipoles).

The behaviour in the centre of mass frame will be similar, except that
the condition for a small interaction will be that both onia should be
unevolved, so 

\begin{equation}
S_{CM} \propto \exp^2 \left[-c\left(\frac{Y - Y_0'}{2}\right)^2
\right] 
= \exp\left[-c\frac{(Y - Y_0')^2}{2} \right]
\end{equation}

\begin{figure}
\begin{center}
\epsfig{file=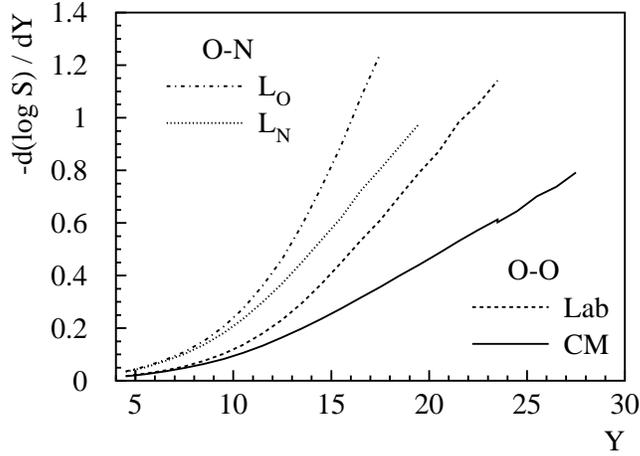, width = 0.6\textwidth}
\caption{The logarithmic derivative of the zero impact parameter
$S$-matrix for centre of mass and lab frames. Curves under the heading
O-O are for onium-onium collisions, while the O-N curves are for
onium-nucleus collisions, where the nucleus consists of two onia with
a random relative transverse position of the order of the onium size.} 
\label{fig:smat_lgdrv}
\end{center}
\end{figure}

\noindent The value of $Y_0'$ is different from $Y_0$ because it
depends on the interaction between the two onia, the details of which
are different in the lab and centre of mass
frames. Figure~\ref{fig:smat_lgdrv} shows results from the Monte Carlo
simulation, plotting the derivative of $\ln S$. The clear straight
line behaviour at the higher values of $Y$ is the signal of the $Y^2$
term in the exponent. Fits to the slopes of the derivatives show
exactly the predicted factor of two between the lab and centre of mass
frames. For $\alps \simeq 0.18$, the value of $c$ is $c \simeq
0.042$. If one performs similar calculations for onium-nucleus
collisions, one expects (and sees) very similar behaviour: in the
$L_N$ frame, the slope is identical to the onium-onium lab frame
results, and the $L_O$ slope is twice the $L_N$ slope (where the
nucleus consists of two onia).

Since, for onium-onium collisions, the asymptotic value of the
$S$-matrix is much smaller in the lab frame than in the centre of mass
frame, one concludes that saturation in the lab frame will increase
the $S$-matrix, and so that the \emph{typical} value of the
$S$-matrix, including saturation will be

\begin{equation}
S^{typical}(Y) \simeq \e^{-c\frac{(Y-Y_0')^2}{2}},
\end{equation}

\noindent very different from the typical value without saturation,
$\sim \exp[-\alps^2\e^{\apm Y}]$. This behaviour of the $S$-matrix
can related to the typical fields in an evolved onium or nucleus
and this is done in \cite{KoMW96}.

\subsection{Phenomenological modification of the dipole branching}

\begin{figure}
\begin{center}
\epsfig{file=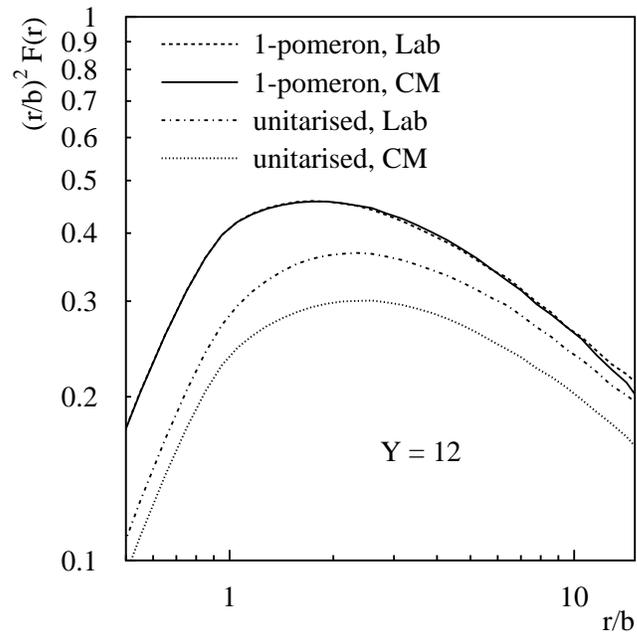, width = 0.6\textwidth}
\caption{The 1-pomeron and unitarised scattering amplitudes for
onium-onium collisions in centre of mass and laboratory frames.}
\label{fig:unit_nosat}
\end{center}
\end{figure}

A second approach to saturation in the real case is intended to give
some information on what happens at non-zero impact parameters. First
one can compare the unitarised amplitude for the lab and centre of
mass frames as a function of $r$ (the amplitudes have been multiplied
by $r^2$ to indicate which regions contribute most to the total cross
section).  This is shown\footnote{Note
that the calculation of unitarisation corrections in the lab frame is
potentially incorrect, because the formula, $f^{(unit)} = 1 -
\exp(-f)$, used for the unitarised interaction from a pair of
configurations which give an interaction $f$ with one pomeron
exchange, breaks down when the number of interacting dipoles in either
of the onia is small and also when the fields in either onium become
too strong.} in figure~\ref{fig:unit_nosat}.  The slight mismatch
between the 1-pomeron curves at moderate $r$ is due to statistical
error (the amplitude at larger values of $r$ comes from very rare
configurations). As one expects, due to the neglect of saturation, the
unitarity corrections in the lab frame are significantly smaller than
in the centre of mass frame. Note in particular though that whereas at
small impact parameters the difference between the two frames is only
a small fraction of the unitarisation corrections, at large impact
parameters, most of the unitarity corrections present in the centre of
mass frame are absent in the lab frame. This suggests that in the lab
frame, saturation effects will in fact be considerably larger than the
unitarity corrections at large $r$. One way of investigating this
further is to try to include some simulation of the effects of
saturation.

The problem is that one doesn't know how to include saturation
effects. Our approach will be to use a very simple model of saturation
which retains the basic properties of normal dipole evolution and
where saturation changes only the rate of dipole branching, not the
spatial distribution. This is not expected to give accurate
quantitative information on evolution with saturation, only to produce
some understanding of the types of effects which are liable to be
important if one does take saturation into account properly.  In
analogy with the toy model, we will use a rate of branching, $R^s_i$,
of a dipole $i$ of size $b$ into two, of sizes $c_<$ (the smaller) and
$c_>$ (the larger) including saturation, of

\begin{equation}
\frac{R^s_i (b \to c_<, c_>)}{R_i (b \to c_<, c_>)} = 
\left(1 - \e^{-O_i / \Omega}\right). \frac{\Omega}{O_i}
\label{eq:rlmdlstrn}
\end{equation}

\begin{figure}
\begin{center}
\epsfig{file=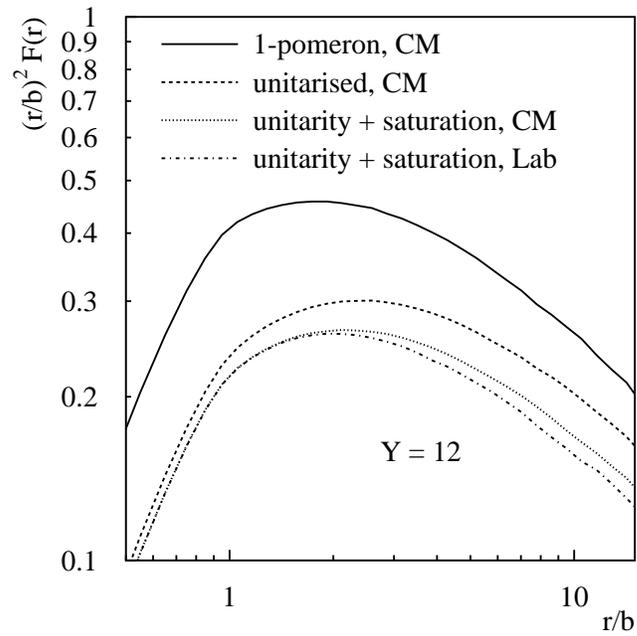, width = 0.6\textwidth}
\caption{The onium-onium scattering amplitude, showing the frame
invariance that results when both unitarity and saturation corrections
are included.}
\label{fig:unit_sat}
\end{center}
\end{figure}

\noindent and where $R_i$ is the normal rate of branching (i.e.\
without saturation). The overlap measure of saturation, $O_i$, is used
because it is much faster to calculate, and during the Monte Carlo
evolution, it will be the calculation of the saturations for each
dipole which will be the limiting factor. Since the functional form of
this saturation is taken from the toy model, one does not know what
normalisation $\Omega$ to use for the overlaps. The procedure used for
choosing $\Omega$ will be to find a value which gives a frame
independent result for $F(r=0)$ ($r$ is the relative impact parameter
between the colliding onia) at a single value of rapidity (which is
always possible), and then to see whether this also works at other
values of $r$ and rapidity. This has been done for $Y=8$, where we
take $\Omega \simeq 0.33$. Reexpressing this in terms of the a more
natural quantity, the probe interaction $I$, this corresponds to a
modification of the branching rate by a factor $\simeq (1 + I/1.3)$,
which seems reasonable.

Figure~\ref{fig:unit_sat} shows the scattering amplitudes for centre
of mass and lab frames, including saturation for $Y=12$. The
one-pomeron and unitarised centre of mass results are also plotted for
reference. The procedure has worked insofar as the results in the two
frames agree for small $r$ for the second rapidity. At larger $r$
there is still a small discrepancy between the two frames: the lab
frame result is too low --- there is a little bit too much saturation
at large $r$ --- this discrepancy is however small compared to the
overall effect of saturation at those values of $r$. The same is found
at all other values of rapidity (including $Y=8$). Other
implementations of saturation (i.e.\ with different forms for
eq.~(\ref{eq:rlmdlstrn})), give very similar results, as long as the
effect of saturation on branching rates is linear for small overlaps.

\begin{figure}
\begin{center}
\epsfig{file=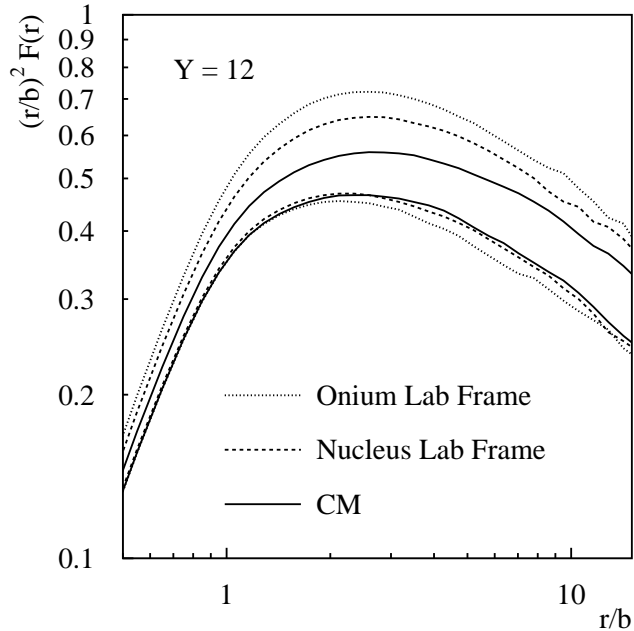, width = 0.6\textwidth}
\caption{Onium-nucleus collisions (nucleus consists of two onia) in
various frames: the upper set of three curves contains only unitarity
corrections, while the lower set has both unitarity and saturation
corrections.} 
\label{fig:us_nclr}
\end{center}
\end{figure}

As a second check on this form of saturation, one can examine
onium-nucleus collisions. As well as the two lab frames discussed
earlier we examine also a ``centre of mass'' frame (or more accurately, a
frame in which the mean number of central dipoles is the same for the
onium and the nucleus). Neglecting saturation one expects the $L_O$
frame to have the lowest unitarity corrections (since with the nucleus
containing all the rapidity, it will have the largest saturation
corrections), followed by the $L_N$ and CM
frames. Figure~\ref{fig:us_nclr} shows this ordering of the
amplitudes. Once saturation is included (using exactly the same
procedure as for onium-onium collisions), the three frames give very
similar amplitudes, though again there is a small discrepancy at
larger $r$. 

It would be interesting to examine how this implementation of saturation
affects the typical $S$-matrix for small impact parameters, and
whether this is in accord with the arguments presented
earlier. However the asymptotic $e^{-c(Y-Y_0)^2/2}$ behaviour sets in
only for relatively large rapidities (one needs $Y\ge 14$), and
calculation of the saturation corrections at these rapidities is
extremely time consuming (because the calculation time is proportional
to the square of the number of dipoles) and therefore not feasible.

Considering the difference between the amplitude in the centre of mass
frame with and without saturation, at small $r$, saturation seems to
have a fairly small effect on the centre of mass amplitude. This is as
expected from the basic argument that each onium has half the
rapidity, and should be well away from saturation when unitarity
corrections start to set in.  At large $r$, however, the effect of
saturation is bigger, and in fact of the same order as the
unitarisation correction (which relates to the observation for
figure~\ref{fig:unit_nosat} that at larger $r$ there is a big
difference between the centre of mass and lab frame results,
indicating that saturation must have a significant effect at large
$r$).

\begin{figure}
\begin{center}
\epsfig{file=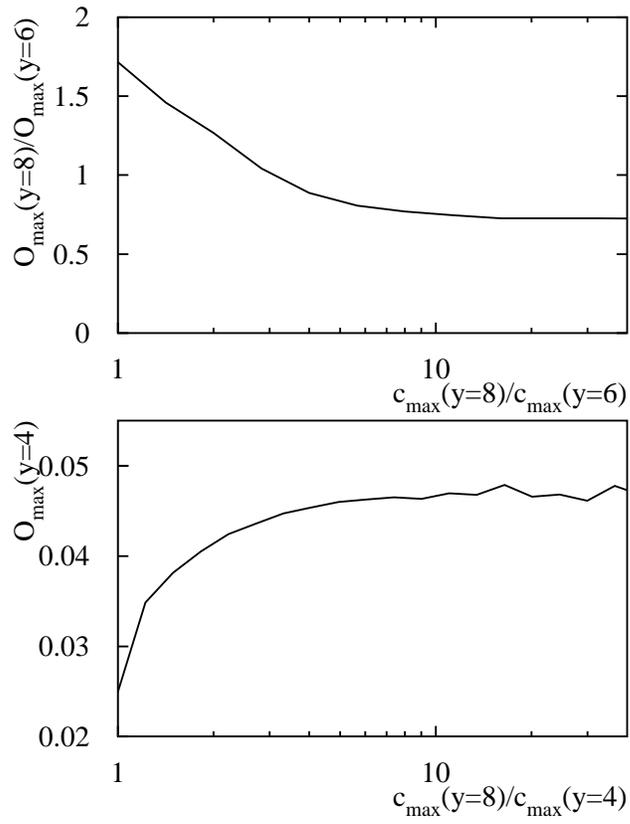, width = 0.6\textwidth}
\caption{Saturation at different stages of evolution. The maximum
dipole size at a given rapidity is $c_{max}$, and $O_{max}$ is the
maximum overlap of any dipole between $c_{max}$ and $c_{max}/3$
(essentially a measure of the saturation on the scale of
$c_{max}$). Errors are large at large $r$ because of a lack of
statistics in that region.}
\label{fig:strn_pths}
\end{center}
\end{figure}

The dynamics involved in producing the effects saturation at large $r$
are complex: because the unitarisation corrections are small, the
densities of dipoles must be small, and one might be misled into
thinking that saturation corrections should unimportant. One reason
why this argument fails is that there are a variety of ways in which
large $r$ interaction can occur. A large dipole can be produced early
in the evolution, which can then evolve to produce many other dipoles
of similar size and the final density of dipoles will be indicative of
the relevant saturation corrections (these evolution paths tend to
also have significant unitarity corrections). Another evolution path
involves only small dipoles for most of the rapidity and a branching
to larger dipoles at a late stage. The evolution of the small dipoles
may have been subjected to significant saturation; but once the
branching occurs to large dipoles, unitarity corrections and any
further saturation corrections (at the large scale) are no longer
affected by what is happening on the small scale. The history of
saturation can therefore have an important effect on the evolution,
even though the diluteness of the final configuration might make one
think one could neglect saturation. This is illustrated in the upper
part of figure~\ref{fig:strn_pths}: at two values of rapidity ($y=6$
and $y=8$), the maximum dipole size, $c_{max}(y)$ is determined. The
maximum overlap, $O_{max}$ is then calculated for all dipoles with
size $c$, where $c_{max} > c > c_{max}/3$, in effect giving the
saturation on the scale of the largest dipole. The figure shows the
ratio $O_{max}(y=8)/O_{max}(y=6)$, the increase in saturation (on the
largest scale), as a function of $c_{max}(y=8)/c_{max}(y=6)$, the
increase in the largest dipole size. The main point is that when there
is a large increase in dipole size, the ratio of the saturations drops
to below one, confirming that if one looks only at the final
configuration, the history of the saturation can be hidden. A similar
reduction of saturation with increasing rapidity can probably also
take place for the small dipoles produced from large ones (since the
small ones will be more dilutely spread out).

A second effect is shown in the lower part of
figure~\ref{fig:strn_pths}: the production of a large dipole tends to
be a rare event. By a simple probabilistic argument, the larger
the number of dipoles on a small scale, the greater the likelihood
that at least one of them will branch to give a large dipole, which
means that evolution paths that produce a large dipole are more likely
to include an intermediate high saturation stage. The lower plot of
figure~\ref{fig:strn_pths} demonstrates this by showing the maximum
saturation (again on the largest scales) at $y=4$ as a function of the
ratio of the largest scales at $y=8$ and $y=4$. As the ratio of the
scales increases so does the maximum intermediate saturation.

Finally in the centre of mass frame, one has the possibility of
situations where one of the onia is small and moderately dense, while
the other has a dilute distribution of small dipoles (the only ones
which will interact with the small dense onium) at large $r$. The
saturation corrections in the evolution of the small dense onium will
then be of the same order as the unitarity corrections between the two
at large $r$ onia, i.e.\ there is an asymmetry between the two onia
(and arguments for neglecting saturation when unitarity corrections
set in, rely on the symmetry between the two onia). This is distinct
from the $r=0$ case where moderate densities of dipoles in \emph{each}
onium led to strongly unitarised amplitude.

The conclusions therefore, from the simple simulation of saturation
effects, and from the other arguments presented above, is that at
large impact parameters, it is not necessarily safe to neglect
saturation corrections compared to unitarity corrections, and further
that large dipoles, even when dilute, may have been produced via a
high saturation stage. This might be of some concern when determining
the corrections to the total cross section, which has a significant
component coming from moderate to large $r$. However, to the extent
that it is possible to gain any quantitative information from our
simple simulation of saturation, one finds that in the total cross
section, saturation corrections (up to $Y\simeq 16$) are no more than
a third of the unitarity corrections.

\section*{Acknowledgements}
We are grateful to B.R.~Webber for many useful discussions. One of the
authors (AM) wishes to thank Peter Landshoff and Bryan Webber for
their kind hospitality and PPARC for a Visiting Fellowship during his
visit to Cambridge, in June of 1995, where this work began.

\section*{Appendix}
This section demonstrates the invariance of the one-pomeron scattering
amplitude on the choice of frame. We show that it is invariant for a
single dipole in each onium. Because each dipole evolves
independently, and because its interaction with the other onium is
independent of the other dipoles in the onium to which it belongs,
this will imply that the onium-onium scattering amplitude is invariant
under changes of frames at the one-pomeron level.

Let the first dipole have sources at $b_1$ and $b_2$, and the second
dipole have sources at $b_3$ and $b_4$ (with the $b_n$ being complex
variables, and $\bar{b}_n$ their complex conjugates). The condition
for invariance of the scattering amplitude on the division of a small
amount of rapidity between them is, eq.~(\ref{eq:realinv}), which can
be written as

\begin{eqnarray}
\int \df b_5 \df \bar{b}_5 \left| \frac{b_{12}}{b_{15}b_{25}} \right|^2 
\left( \ln^2\left|\frac{b_{13}b_{54}}{b_{14} b_{53}}\right|^2
 +     \ln^2\left|\frac{b_{23}b_{54}}{b_{24} b_{53}}\right|^2
 -     \ln^2\left|\frac{b_{13}b_{24}}{b_{14} b_{23}}\right|^2
\right)
\\ = 
\int \df b_5 \df \bar{b}_5 \left| \frac{b_{34}}{b_{35}b_{45}} \right|^2
\left( \ln^2\left|\frac{b_{13}b_{25}}{b_{15} b_{23}}\right|^2
 +     \ln^2\left|\frac{b_{14}b_{25}}{b_{15} b_{24}}\right|^2
 -     \ln^2\left|\frac{b_{13}b_{24}}{b_{14} b_{23}}\right|^2.
\right) \nonumber
\label{eq:onedipinv}
\end{eqnarray}

\noindent To see that this holds one notes that each of the integrands 
(including the $\df b_5 \df \bar{b}_5$ measure) is invariant under
conformal transformations. The particular conformal transformation 

\begin{equation}
b \to \frac{b + A}{Cb - 1}
\end{equation}

\noindent with 

\begin{eqnarray}
A & = & \frac{b_1 - b_2 + b_3 - b_4}{b_1 b_3 - b_2 b_4} \\ \nonumber
C & = & \frac{b_1^{-1} - b_2^{-1} + b_3^{-1} - b_4^{-1}}
{b_2^{-1} b_4^{-1} - b_1^{-1} b_3^{-1}}
\end{eqnarray}

\noindent has the property of mapping $b_1 \to b_3$, $b_2 \to b_4$,
$b_3 \to b_1$ and $b_4 \to b_2$, thus converting the lower integral in
eq.~(\ref{eq:onedipinv}) to the upper integral (and vice-versa). Since
the integrands are invariant under conformal transformations, this
proves the equality, and therefore the invariance of the one-pomeron
onium-onium scattering amplitude on the frame in which the calculation
is performed, if the evolution is carried out using the dipole kernel
and the interaction is by two gluon exchange.

%\bibliographystyle{aip}
%\bibliography{jouabrv,bib}

\begin{thebibliography}{10}

\bibitem{BaLi78}
Y.~Y. Balitski\v{\i} and L.~N. Lipatov,
\newblock Sov. Phys. JETP  28 (1978) 822.

\bibitem{KuLF77}
E.~A. Kuraev, L.~N. Lipatov, and V.~S. Fadin,
\newblock Sov. Phys. JETP  45 (1977) 199.

\bibitem{Lipa86}
L.~N. Lipatov,
\newblock Sov. Phys. JETP  63 (1986) 904.

\bibitem{Doks77}
Y.~L. Dokshitzer,
\newblock Sov. Phys. JETP  46 (1977) 641.

\bibitem{GrLi72}
V.~N. Gribov and L.~N. Lipatov,
\newblock Sov. J. Nucl. Phys  15 (1972) 78.

\bibitem{AlPa77}
G.~Altarelli and G.~Parisi,
\newblock Nucl. Phys.  B 126 (1977) 298.

\bibitem{MuNa87}
A.~H. Mueller and H.~Navelet,
\newblock Nucl. Phys.  B 282 (1987) 727.

\bibitem{FaLi96}
V.S.Fadin and L.N.Lipatov,
\newblock DESY preprint 96-020, hep-ph/9602287  (1996).

\bibitem{Muel94a}
A.~H. Mueller,
\newblock Nucl. Phys.  B 415 (1994) 373.

\bibitem{MuPa94}
A.~H. Mueller and B.~Patel,
\newblock Nucl. Phys.  B 425 (1994) 471.

\bibitem{Muel94b}
A.~H. Mueller,
\newblock Nucl. Phys.  B 437 (1995) 107.

\bibitem{Muel95}
Z.~Chen and A.~H. Mueller,
\newblock Nucl. Phys. B 451 (1995) 579.

\bibitem{NiZZ94a}
N.~N. Nikolaev, B.~G. Zakharov, and V.~R. Zoller,
\newblock JETP Lett. 59 (1994) 6.

\bibitem{Sala95b}
G.~P. Salam,
\newblock Nucl. Phys.  B 461 (1996) 512.

\bibitem{AbGK74}
V.~Abramovskii, V.~Gribov, and O.~Kancheli,
\newblock Sov. J. Nucl. Phys  18 (1974) 308.

\bibitem{KoNO72}
Z.~Koba, H.~B. Nielsen, and P.~Olesen,
\newblock Nucl. Phys.  40 (1972) 317.

\bibitem{Sala96a}
G.~P. Salam,
\newblock Cavendish-HEP preprint 95/07, hep-ph/9601220 (1996).

\bibitem{GrLR83}
L.~V. Gribov, E.~M. Levin, and M.~G. Ryskin,
\newblock Phys. Rep. 100 (1983) 1.

\bibitem{KoMu75}
J.~Koplik and A.~H. Mueller,
\newblock Phys. Rev.  12 (1975) 3638.

\bibitem{KoMW96}
Y.~Kovchegov, A.~H. Mueller, and S.~Wallon,
\newblock in preparation.

\end{thebibliography}

\end{document}